\documentclass[useAMS,usenatbib]{mn2e}

\usepackage{graphicx}

\def\ms{\mbox{$M_\odot$}}
\def\ls{\mbox{$L_\odot$}}

\title[The formation of NGC\,6868 and NGC\,5903]{Star formation, metallicity gradient and 
ionized gas: clues to the formation of the elliptical galaxies NGC\,6868 and NGC\,5903}

\author[M.G. Rickes et al.]{ M.G. Rickes$^1$, M.G. Pastoriza$^{1}$ and C. Bonatto$^{1}$\\
$^{1}$Departamento de Astronomia, Universidade Federal do Rio Grande do Sul,
Av. Bento Gon\c calves 9500, Porto Alegre, RS, Brazil. \\
}

\begin{document}

\label{firstpage}

\maketitle

\begin{abstract}
The stellar population, metallicity distribution and ionized gas in the elliptical galaxies
NGC\,6868 and NGC\,5903 are investigated in this paper by means of long-slit spectroscopy and
stellar population synthesis. Lick indices in both galaxies present a negative gradient indicating
an overabundance of Fe, Mg, Na and TiO in the central parts with respect to the external regions.
We found that $Mg_2$ correlates both with ${\rm FeI}_{\lambda5270}$ and ${\rm FeI}_{\lambda5335}$,
suggesting that these elements probably underwent the same enrichment process in NGC\,6868. However,
only a marginal correlation of $Mg_2$ and ${\rm FeI}_{\lambda5270}$ occurs in NGC\,5903. The lack of
correlation between computed galaxy mass and the $Mg_2$ gradient suggests that these elliptical galaxies
were formed by merger events. The stellar population synthesis shows the presence of at least two
populations with ages of  13\,Gyr and 5\,Gyr old in both galaxies.

We have estimated the metallicity of the galaxies using SSP models. The central region of NGC\,6868 
($|R|\la0.5$\,kpc) presents a deficiency of alpha elements with respect to iron and solar metallicity. 
The external parts,  present a roughly uniform distribution of $\rm [\alpha/Fe]$ ratios and metallicities 
ranging from $[Z/Z_\odot] = -0.33$ and solar. A similar conclusion applies to NGC\,5903.

Concerning the emitting gas conspicuously detected in NGC\,6868, we test three hypotheses as ionizing
source: an H\,II region, post-AGB stars and an Active Galactic Nucleus (AGN). Diagnostic diagrams involving
the ratios $[NII]_{\lambda6584}/H\alpha$, $[OI]_{\lambda6300}/H\alpha$ and $[SII]_{\lambda6717,31}/H\alpha$,
indicate that values measured in the central region of NGC\,6868 are typical of LINERs. Together with
the stellar population synthesis, this result suggests that the main source of gas ionization in NGC\,6868
is non-thermal, produced by a low-luminosity AGN, probably with some contribution of shocks to explain
ionization at distances of $\sim3.5$\,kpc from the nucleus.

\end{abstract}

\begin{keywords}
 Elliptical galaxies: NGC\,6868 and NGC\,5903 - fundamentals parameters
\end{keywords}

\section{Introduction}
\label{Intro}

A question still open to debate is how elliptical galaxies are formed. Three scenarios have
been proposed to address this issue: {\em (i)} monolithic collapse of gas clouds linked to high
energy dissipation rates \citep{arim87}; {\em (ii)} galaxy merger \citep{burk00}; and {\em (iii)}
hybrid models that involve an early monolithic collapse followed by merger events at later times
\citep{ogan05}. A consequence of the monolithic collapse (associated to a large-scale star formation)
is that a conspicuous metallicity gradient is expected to establish along the full range of galactic
radius (e.g. \citealt{car93a}; \citealt{ChioCar02}). However, if merger events are important, the
gradient is expected to disappear. A few ellipticals present strong signs of recent dynamical
perturbation \citep{schw90}, which may provide important clues to their formation processes and
evolution.

Important information on the above issues has been derived from metal line-strength indices
(e.g. \citealt{davis87}) and their radial variation within galaxies. $Mg2$ line-strength
distribution in early-type galaxies, for example, can vary considerably, ranging from essentially
featureless to structured profiles showing, e.g., changes of slope possibly associated with kinematically
decoupled cores, or anomalies in the stellar population of some ellipticals \citep{car93b}.

In addition, the origin of the ionized gas in ellipticals is another controversial issue, because
until recently it was believed that the interstellar medium (ISM) in these galaxies was
not significant. However,  studies by, e.g. \citet{phi86} have shown that more
than 50\% of the elliptical and lenticular galaxies contain significant amounts of
ionized gas. More recently, \citet{macc96} have traced ionized gas in a sample of 73
elliptical and SO galaxies by means of narrow-band filters centered on ${\rm H}\alpha$
and [NII]. They detected emission gas in about 3/4 of the galaxies, with masses
amounting to $10^3$ to $10^5\,\ms$. 

In the present work we investigate properties of the elliptical galaxies NGC\,6868 and NGC\,5903,
because previous studies have shown that they contain important interstellar medium (\citealt{macc96}; 
\citealt{ferr99}), have complex stellar and gas kinematics (\citealt{ca00}). Besides, both ellipticals 
belong to galaxy groups, which might imply past close encounters (and even mergers)
with companions. Potential consequences of such past interactions are the triggering of discrete star
formation events with stellar populations differing in metallicity and age (\citealt{wor92}) and the
presence of large-scale dynamical perturbations that may affect galaxy morphology by building subsystems
such as a stellar and/or dust disk (\citealt{ca00}). In this context, NGC\,6868 and NGC\,5903 can be taken
as interesting candidates where the stellar population and metallicity distribution can be investigated,
which are important parameters to understand the formation and evolution of these galaxies.

The main goal of the present paper is to investigate, by means of long-slit spectroscopy, the origin,
star-formation history, metallicity and ionized gas distribution in NGC\,6868 and NGC\,5903. The analyses
will be based mostly on a stellar population synthesis method (based on templates built with star
clusters and H\,II regions), and the radial distribution of absorption and emission lines.

This paper is organized as follows. In Sect.~\ref{Observations} we describe the observations. Lick indices
are measuremed are discussed in Sect.~\ref{Lick}. The stellar population synthesis and results are described
in Sect.~\ref{StellarPop}. Metallicity and ionized gas are discussed in Sect.~\ref{MetGas}. Discussions are 
in Sect.~\ref{Dicussion}.

\section{The Target Galaxies and Spectroscopic observations}
\label{Observations}

\begin{figure}
\resizebox{\hsize}{!}{\includegraphics[angle=0]{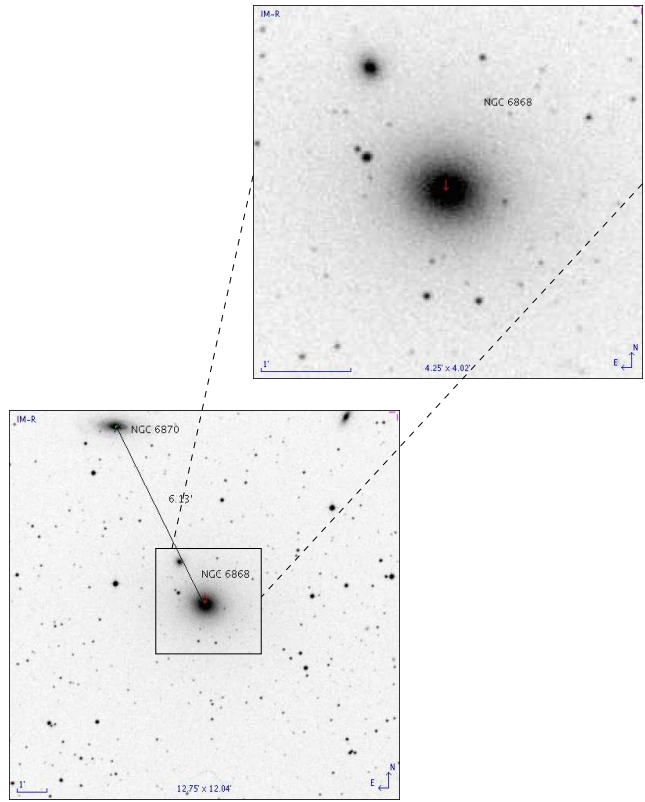}}
\caption{Bottom panel: R image of NGC\,6868 showing a scale of
$12.75\arcmin\times12.04\arcmin$. Top panel: same as before for the region
$4.25\arcmin\times4.02\arcmin$ centered on NGC\,6868.}
\label{figngc6868}
\end{figure}

NGC\,6868 is classified as E3 in RC3\footnote{The Third Reference Catalogue of bright galaxies
(\citealt{vaucouleurs}).}. It is part of the GR28 cluster which contains as well the galaxies NGC\,6861,
NGC\,6870, NGC\,6851 and NGC\,6861D \citep{maia89}. Figure~\ref{figngc6868} shows an R band image of
the field containing NGC\,6868 in two scales (obtained from the NASA/IPAC Extragalactic Database - NED).

The effective radius of NGC\,6868 is $Re=40\arcsec$ \citep{car93a}. NGC\,6868 is a radio source \citep{sava77}
and emits in the infrared as indicated by 60 and $100\,\mu$m IRAS observations. The total luminosity in the
medium infrared is $15.2\times10^8\,\ls$, and the dust mass estimated from grain emission considerations is
$\approx70\,\ms$ \citep{ferr02}.

The velocity field and velocity dispersion profiles are symmetrical about the galaxy center. However,
the gas is not observed to follow regular orbits \citep{zeil96}.  The latter work shows evidence in
several position angles that the rotation curve does not remain flat in the outer parts but bends back
towards a value close to the systemic velocity.

\begin{figure}
\resizebox{\hsize}{!}{\includegraphics[angle=0]{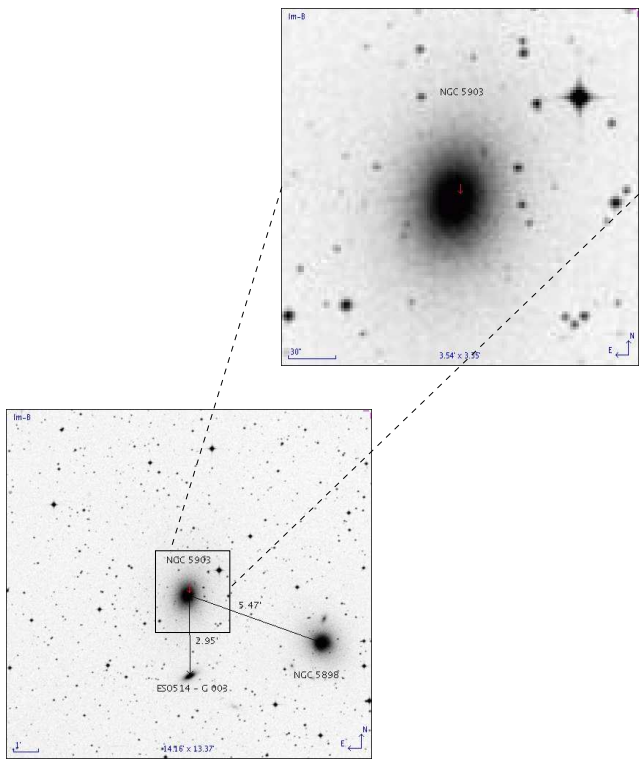}}
\caption{Bottom panel: B image of the field of NGC\,5903 showing a scale of
$14.16\arcmin\times13.37\arcmin$. Top panel: same as before for the region
$3.54\arcmin\times3.35\arcmin$ centered on NGC\,5903.}
\label{figngc5903}
\end{figure}

NGC\,5903 (morphological type E2 - RC3) forms a pair with the elliptical galaxy NGC\,5898; both galaxies
are the brightest members of a small group composed of NGC\,5903, NGC\,5898 and ESO\,514-G\,003 \citep{maia89}.
NGC\,5903 presents a small rotation around  the semi-major axis \citep{spar85}. Dust is observed in a small
region ($\sim100$\,pc) \citep{ferr99}. NGC\,5903 presents X-ray emission with no evidence of ionized gas
\citep{osull01}. The effective radius of NGC\,5903 is $Re=37\arcsec$ \citep{car93a}. B images (from NED) of
the field containing NGC\,5903 are shown in Fig.~\ref{figngc5903}.

NGC\,6868 and NGC\,5903 were observed with the 3.6\,m ESO telescope (in Chile) equipped with EFOSC1.
The observational setup included the grism 0150 for the range 5140 to 6900\,\AA, with a dispersion of
$\rm3.4\,\AA\,pixel^{-1}$, and the CCD\#2 ($512\times512$ Tektronix). This setup was used in order to
maximize the spectral coverage $\Delta\lambda\,5100-6800$\AA\ and obtain the largest possible number
of stellar absorption features and gas emission lines without compromising the spectral resolution of
$\rm3.4\AA\,pixel^{-1}$ \citep{ca00}. The spatial scale of the observational setup was
$\rm0.6\arcsec\,pixel^{-1}$ with a $3.1\arcmin$ long slit; the slit width was fixed at $1.5\arcsec$,
approximately equal to the seeing.

At least two spectra were obtained along the position angle in order to increase the signal-to-noise
ratio and for cosmic ray removal. For NGC\,6868 three long-slit spectra were obtained with 40\,min of
exposure time; the slit was oriented along the semi-major axis ($\rm PA=120^\circ$). Two 40\,min spectra
were obtained for NGC\,5903 with the slit oriented along the semi-major axis ($\rm PA=340^\circ$). The
seeing varied from $ 1.2\arcsec $ to $ 1.8\arcsec$. Several spectrophotometric standard stars were observed
at the beginning and end of each night in order to flux calibrate the spectra. The spectra were processed
with standard IRAF tasks. The mean bias was subtracted from each spectra, which were subsequently divided
by a normalized dome flat-field to remove variations in sensitivity. The centroids of the non-saturated lines
in the He+A comparison spectra were measured in order to obtain the wavelength calibration and to map the
line curvature. Galaxy and standard stars spectra were rectified and rebinned to a logarithmic scale with a
constant step of $\rm85\,km\,s^{-1}$. The sky contribution for each galaxy spectra was determined by taking
the median value between two windows of 20 pixels wide at both ends of the slit and subtracted.

\begin{figure}
\resizebox{\hsize}{!}{\includegraphics[angle=0]{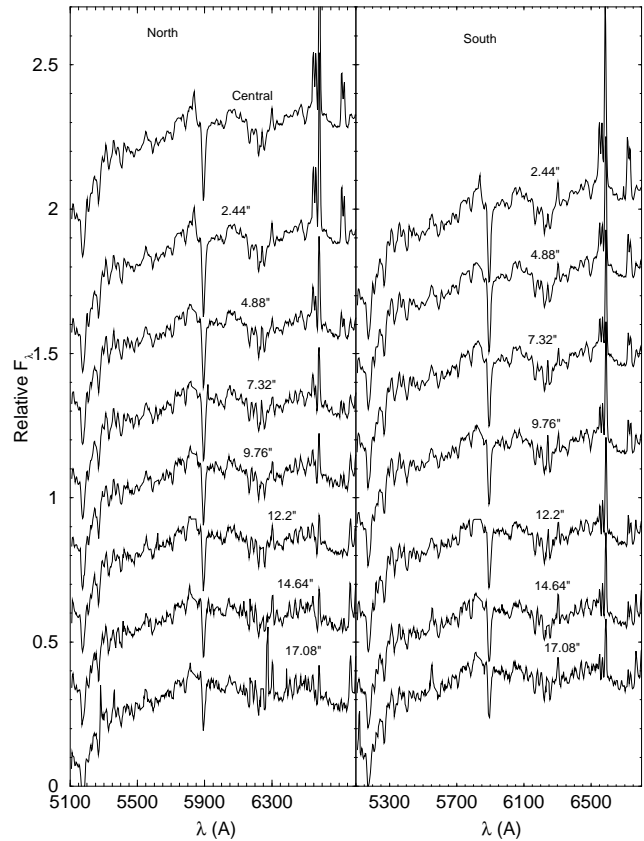}}
\caption{Rest-frame spectra extracted from NGC\,6868, corrected for foreground reddening. The
distance of each extraction to the galaxy center is given in arcsec. North and south extractions 
are shown in the left and right panels, respectively. Arbitrary constants have been added to the 
spectra for clarity. Conspicuous emission lines (especially $H_\alpha$ and [NII]$\lambda\lambda6548,6584$)
occur in all spectra.}
\label{fig3}
\end{figure}

\begin{figure}
\resizebox{\hsize}{!}{\includegraphics[angle=0]{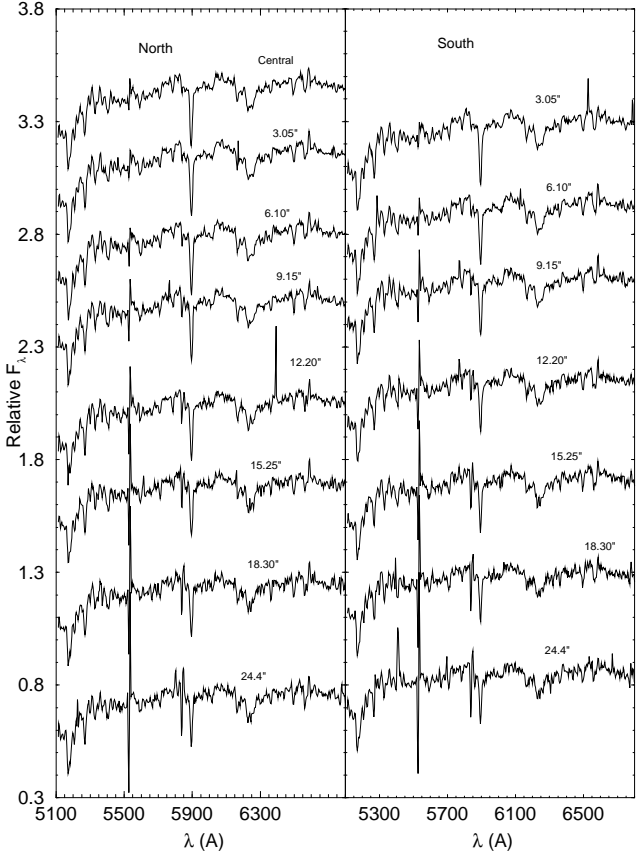}}
\caption{Same as Fig.~\ref{fig3} for NGC\,5903.}
\label{fig4}
\end{figure}

Spectra were extracted in both galaxies for the central and 7 regions symmetrically distributed
along the north-south direction. Pairs of spectra obtained at the same distance of the galaxy center
were averaged together in order to increase the signal to noise ratio. We have estimated in less than
1\% the relative spectrophotometric errors between both extractions at each position. The distance of
each extraction to galaxy center is given in cols.~1 to 3 of Table~\ref{ew6868}, respectively in terms
of angular, effective and absolute radii. For redshift correction we used the conspicuous absorption line
$NaI_{5895}$ as reference. Foreground reddening was corrected assuming the Galactic extinction law
of \citet{CCM89}, with
$A_V=0.085$ (NGC\,6868) and $A_V=0.332$ (NGC\,5903) using the $E(B-V)$ values taken from NED and assuming
$A_V=3E(B-V)$. The spectra of NGC\,6868 and NGC\,5903, corrected for redshift and foreground reddening,
are shown in Figs.~\ref{fig3} and \ref{fig4}, respectively. The distances of NGC\,6868 and NGC\,5903,
computed assuming the Hubble constant $\rm H_O=75\,km\,s^{-1}$, are
38\,Mpc and 34\,Mpc respectively. Accordingly, one arcsec is equivalent to 184\,pc for NGC\,6868 and
164\,pc for NGC\,5903.

\section{Lick indices}
\label{Lick}

NGC\,6868 and NGC\,5903 present strong absorption lines of neutral iron and sodium, as well as Mg2 and 
TiO bands in their visible spectra (Figs.~\ref{fig3} and \ref{fig4}) that, together with the continuum
distribution, can be used to estimate the age and metallicity of the stellar population. In the present
work we adopt the Lick system \citep{fab85,wor94} to measure equivalent widths (EWs) for the absorption
features Mg$_{2}~\lambda5176$, ${\rm FeI}_{\lambda5270}$, ${\rm FeI}_{\lambda5335}$, ${\rm FeI}_{\lambda5406}$,
${\rm FeI}_{\lambda5709}$, ${\rm FeI}_{\lambda5782}$,  ${\rm NaI}_{\lambda5895}$ and ${\rm Tio}_{\lambda6237}$.
Continuum values, normalized at $\lambda5870$\AA, were measured at $\lambda5300, 5546, 5650, 5800, 6173,
6620$, and $6640~$\AA. The velocity dispersion in NGC\,6868 varies from $\rm\sim260\pm20\,km\,s^{-1}$ in
the center to $\rm\sim220\pm30\,km\,s^{-1}$ in the external regions and from $\rm\sim240\pm20\,km\,s^{-1}$
in the center to $\rm\sim200\pm40\,km\,s^{-1}$ in the external regions of NGC\,5903 (\citealt{ca00}).
Therefore, the observed EW values were corrected for line broadening due to stellar velocity dispersion.
The spectrum of the G giant star HR\,5333, assumed to have zero intrinsic velocity, was broadened with a
series of Gaussian filters with velocity dispersion $\sigma$ varying from 0 to $\rm300\,km\,s^{-1}$ in steps
of $\rm10\,km\,s^{-1}$.

For each absorption feature considered we calculate an empirical correction index ${\rm C(\sigma)}$, 
so that $C(\sigma)_{Mg2}=\frac{Mg2(0)}{Mg2(\sigma)}$, and $C(\sigma)_{FeI} =\frac{W_\lambda(0)}
{W_\lambda(\sigma)}$. We calculate the correction index $C(\sigma)$ for each extracted spectrum using
the $\sigma$ values given by \citet{ca00}. The resulting correction factors, computed for each feature
in all extractions, for both galaxies are given in Table~\ref{tabCC}. EWs measured in the spectra of
NGC\,6868 and NGC\,5903 are given in Table~\ref{ew6868}. Except for EW(Mg$_{2}~\lambda5176$), which is
measured in magnitude, the remaining EWs are given in \AA\ (e.g. \citealt{rickes04}). Different values
of EWs are computed allowing for three choices of the continuum level around a given feature, which
basically reflect the subjectiveness associated to continuum determination. EWs given in Table~\ref{ew6868}
correspond to the average values obtained from this process; the uncertainties correspond to the respective
standard deviation.

\begin{table}
\renewcommand{\tabcolsep}{1.0mm}
\renewcommand{\arraystretch}{1.1}
\caption{Correction factors $C(\sigma)$}
\label{tabCC}
\begin{tabular} {l|cccccccc}
\hline
\hline
\multicolumn{9}{c}{NGC\,6868}\\
\hline
$\lambda$(\AA) &$0\arcsec$&$2.4\arcsec$&$4.8\arcsec$&$7.3\arcsec$&$9.7\arcsec$&$12.2\arcsec$&$14.6\arcsec$&$17.08\arcsec$ \\
\hline
5176            & $1.21$    & $1.22$ & $1.21$  & $1.21$  & $1.21$  & $1.21$  & $1.21$  & $1.24$  \\
5270            & $1.10$    & $1.12$ & $1.11$  & $1.11$  & $1.11$  & $1.11$  & $1.10$  & $1.13$  \\
5335            & $1.22$    & $1.26$ & $1.23$  & $1.23$  & $1.23$  & $1.23$  & $1.26$  & $1.27$  \\
5406            & $1.20$    & $1.23$ & $1.24$  & $1.24$  & $1.24$  & $1.24$  & $1.23$  & $1.24$  \\
5709            & $1.13$    & $1.16$ & $1.15$  & $1.15$  & $1.15$  & $1.15$  & $1.16$  & $1.18$  \\
5782            & $1.18$    & $1.18$ & $1.18$  & $1.18$  & $1.18$  & $1.18$  & $1.18$  & $1.25$  \\
5895            & $1.09$    & $1.10$ & $1.09$  & $1.10$  & $1.10$  & $1.10$  & $1.10$  & $1.09$  \\
6237            & $1.08$    & $1.08$ & $1.08$  & $1.08$  & $1.08$  & $1.08$  & $1.08$  & $1.09$  \\
\hline
\multicolumn{9}{c}{NGC\,5903}\\
\hline
$\lambda$(\AA)&$0\arcsec$&$3.05\arcsec$&$6.10\arcsec$&$9.15\arcsec$&$12.20\arcsec$&$15.25\arcsec$&$18.30\arcsec$& $24.40\arcsec$ \\
\hline
5176      & $1.07$ & $1.08$  & $1.06$  & $1.05$  & $1.05 $  & $1.05 $  & $1.05 $  & $1.08 $  \\
5270      & $1.10$ & $1.09$  & $1.07$  & $1.08$  & $1.08 $  & $1.08 $  & $1.08 $  & $1.09 $  \\
5335      & $1.26$ & $1.25$  & $1.21$  & $1.21$  & $1.21 $  & $1.21 $  & $1.21 $  & $1.25 $  \\
5406      & $1.36$ & $1.29$  & $1.30$  & $1.30$  & $1.30 $  & $1.30 $  & $1.30 $  & $1.29 $  \\
5709      & $1.40$ & $1.35$  & $1.37$  & $1.37$  & $1.37 $  & $1.37 $  & $1.37 $  & $1.35 $  \\
5782      & $1.45$ & $1.45$  & $1.32$  & $1.28$  & $1.28 $  & $1.28 $  & $1.28 $  & $1.45 $  \\
5895      & $1.06$ & $1.04$  & $1.03$  & $1.03$  & $1.03 $  & $1.03 $  & $1.03 $  & $1.03 $  \\
6237      & $1.14$ & $1.13$  & $1.12$  & $1.12$  & $1.12 $  & $1.12 $  & $1.12 $  & $1.13 $  \\
\hline
\end{tabular}
\end{table}

\begin{table*}
\renewcommand{\tabcolsep}{1.7mm}
\renewcommand{\arraystretch}{1.1}
\footnotesize
\caption{Equivalent widths measured in the spectra of NGC\,6868 and NGC\,5903}
\label{ew6868}
\begin{tabular} {lcccccccccc}
\hline
\multicolumn{11}{c}{NGC\,6868}\\
\hline
$R(\arcsec)$&$R/R_e$&$R(kpc)$&EW (mag)&\multicolumn{7}{c}{EW (\AA)}\\
\cline{5-11}
 &  & &  $\rm Mg_2$5176    &   FeI5270     &    FeI5335   &  FeI5406     &    FeI5709   &    FeI5782   
 &   NaI5895    & TiO6237   \\
\hline
0.00 &  0.00&0.00 & $ 0.27\pm0.01$ & $3.71\pm0.09$ & $3.34\pm0.05$& $2.54\pm0.01$& $1.23\pm0.02$& $1.26\pm0.02$& $5.51\pm0.02$& $5.78\pm0.13$\\
2.44S&  0.06&0.46 & $ 0.25\pm0.01$ & $3.53\pm0.38$ & $3.12\pm0.16$& $2.36\pm0.12$& $1.20\pm0.05$& $1.29\pm0.03$& $5.43\pm0.02$& $5.78\pm0.12$\\
4.88S&  0.12&0.92 & $ 0.23\pm0.01$ & $3.29\pm0.23$ & $3.21\pm0.41$& $2.52\pm0.04$& $1.19\pm0.06$& $1.07\pm0.02$& $4.68\pm0.04$& $4.52\pm0.15$\\
7.32S&  0.18&1.38 & $ 0.23\pm0.01$ & $3.57\pm0.17$ & $3.08\pm0.07$& $2.35\pm0.10$& $1.15\pm0.04$& $0.98\pm0.02$& $4.54\pm0.03$& $4.68\pm0.17$\\
9.76S&  0.24&1.84 & $ 0.22\pm0.01$ & $3.16\pm0.13$ & $2.49\pm0.18$& $2.14\pm0.11$& $1.01\pm0.07$& $0.84\pm0.05$& $4.04\pm0.03$& $4.00\pm0.20$\\
12.2S&  0.31&2.30 & $ 0.21\pm0.01$ & $3.31\pm0.28$ & $2.42\pm0.07$& $1.57\pm0.05$& $1.00\pm0.09$& $0.93\pm0.02$& $4.01\pm0.09$& $3.85\pm0.18$\\
14.64S& 0.37&2.76 & $ 0.18\pm0.01$ & $1.99\pm0.06$ & $2.36\pm0.19$& $1.57\pm0.05$& $0.90\pm0.12$& $0.86\pm0.05$& $3.92\pm0.10$& $4.43\pm0.23$\\
17.08S& 0.43&3.22 & $ 0.20\pm0.01$ & $2.17\pm0.19$ & $2.10\pm0.17$& $1.56\pm0.04$& $0.95\pm0.01$& $0.89\pm0.01$& $3.52\pm0.10$& $3.51\pm0.25$\\
\hline
2.44N&  0.06&0.46 & $ 0.26\pm0.01$ & $3.61\pm0.13$ & $3.01\pm0.07$& $2.41\pm0.16$& $1.15\pm0.03$& $1.10\pm0.01$& $5.48\pm0.01$& $5.77\pm0.15$\\ 
4.88N&  0.12&0.92  & $ 0.22\pm0.01$ & $3.51\pm0.16$ & $3.06\pm0.06$& $2.36\pm0.17$& $1.15\pm0.03$& $0.92\pm0.03$& $4.88\pm0.03$& $5.58\pm0.16$\\ 
7.32N&  0.18&1.38  & $ 0.22\pm0.01$ & $3.23\pm0.20$ & $2.73\pm0.11$& $2.48\pm0.05$& $0.81\pm0.01$& $1.02\pm0.08$& $4.59\pm0.05$& $5.28\pm0.19$\\ 
9.76N&  0.24&1.84 & $ 0.21\pm0.01$ & $3.13\pm0.09$ & $2.68\pm0.11$& $2.43\pm0.13$& $1.04\pm0.04$& $0.97\pm0.02$& $4.27\pm0.03$& $4.39\pm0.14$\\ 
12.2N&  0.31&2.30 & $ 0.21\pm0.01$ & $3.01\pm0.26$ & $2.85\pm0.04$& $2.21\pm0.09$& $1.02\pm0.04$& $0.90\pm0.07$& $4.17\pm0.09$& $4.77\pm0.23$\\ 
14.64N& 0.37&2.76  & $ 0.18\pm0.01$ & $2.94\pm0.21$ & $2.81\pm0.03$& $2.41\pm0.16$& $0.92\pm0.04$& $1.10\pm0.13$& $4.10\pm0.12$& $4.45\pm0.23$\\ 
17.08N& 0.43&3.22 & $ 0.17\pm0.01$ & $2.60\pm0.15$ & $2.36\pm0.16$& $1.77\pm0.13$& $0.77\pm0.02$& $0.99\pm0.02$& $3.72\pm0.11$& $4.37\pm0.22$\\ 
\hline
\multicolumn{11}{c}{NGC\,5903}\\
\hline
0.0 & 0.00&0.00 & $0.32\pm0.01$ & $3.59\pm0.41$ & $3.71\pm0.01$ & $2.17\pm0.02$ & $1.37\pm0.01$ & $1.22\pm0.02$ & $5.08\pm0.28$ & $7.70\pm0.10$ \\
3.05 S & 0.08&0.48 & $0.31\pm0.01$ & $3.85\pm0.15$ & $3.66\pm0.03$ & $2.37\pm0.01$ & $1.08\pm0.03$ & $1.17\pm0.04$ & $4.84\pm0.06$ & $5.57\pm0.10$ \\
6.10 S & 0.16&0.96 & $0.32\pm0.01$ & $4.30\pm0.59$ & $3.02\pm0.02$ & $2.33\pm0.02$ & $1.35\pm0.02$ & $1.19\pm0.02$ & $4.21\pm0.05$ & $5.24\pm0.05$ \\
9.15 S & 0.25&1.44 & $0.31\pm0.01$ & $3.68\pm0.01$ & $3.40\pm0.03$ & $2.69\pm0.05$ & $1.23\pm0.02$ & $1.23\pm0.05$ & $3.88\pm0.11$ & $4.73\pm0.07$ \\
12.20S & 0.33&1.92 & $0.27\pm0.01$ & $3.44\pm0.11$ & $3.47\pm0.10$ & $2.10\pm0.03$ & $1.41\pm0.03$ & $1.13\pm0.04$ & $3.97\pm0.14$ & $5.61\pm0.04$ \\ 
15.25S & 0.41&2.40 & $0.26\pm0.02$ & $2.94\pm0.12$ & $3.13\pm0.05$ & $2.09\pm0.03$ & $1.33\pm0.01$ & $1.13\pm0.02$ & $3.70\pm0.10$ & $5.34\pm0.08$ \\
18.30S & 0.49&2.88 & $0.29\pm0.01$ & $3.26\pm0.14$ & $3.02\pm0.07$ & $1.74\pm0.12$ & $1.30\pm0.04$ & $0.97\pm0.03$ & $3.31\pm0.13$ & $4.70\pm0.08$ \\
24.40S & 0.66&3.90 & $0.28\pm0.02$ & $2.63\pm0.50$ & $3.09\pm0.04$ & $2.08\pm0.13$ & $1.35\pm0.07$ & $0.95\pm0.02$ & $3.41\pm0.15$ & $5.50\pm0.14$ \\
\hline       
3.05 N & 0.08&0.48  & $0.31\pm0.01$ & $3.45\pm0.23$ & $2.40\pm0.06$ & $1.78\pm0.02$ & $1.09\pm0.03$ & $1.09\pm0.03$ & $4.96\pm0.10$ & $7.14\pm0.09$ \\ 
6.10 N & 0.16&0.96 & $0.30\pm0.01$ & $3.39\pm0.10$ & $3.35\pm0.04$ & $2.16\pm0.03$ & $1.23\pm0.02$ & $0.91\pm0.03$ & $4.60\pm0.07$ & $6.54\pm0.10$ \\
9.15 N & 0.25&1.44 & $0.29\pm0.02$ & $3.62\pm0.09$ & $2.99\pm0.05$ & $2.60\pm0.01$ & $1.11\pm0.02$ & $0.86\pm0.04$ & $4.24\pm0.08$ & $6.08\pm0.07$ \\ 
12.20N & 0.33&1.92 & $0.25\pm0.02$ & $3.59\pm0.10$ & $3.03\pm0.06$ & $2.03\pm0.03$ & $0.95\pm0.05$ & $0.89\pm0.02$ & $4.09\pm0.12$ & $4.97\pm0.09$ \\ 
15.25N & 0.41&2.40 & $0.28\pm0.02$ & $3.45\pm0.07$ & $2.74\pm0.03$ & $2.49\pm0.02$ & $1.27\pm0.03$ & $0.91\pm0.05$ & $4.12\pm0.15$ & $5.99\pm0.05$ \\
18.30N & 0.49&2.88 & $0.27\pm0.01$ & $3.39\pm0.09$ & $3.37\pm0.15$ & $2.23\pm0.03$ & $1.34\pm0.02$ & $1.01\pm0.02$ & $3.76\pm0.07$ & $4.90\pm0.08$ \\ 
24.40N & 0.66&3.90 & $0.30\pm0.02$ & $3.28\pm0.06$ & $3.33\pm0.06$ & $1.98\pm0.03$ & $1.09\pm0.02$ & $0.72\pm0.10$ & $3.35\pm0.08$ & $5.48\pm0.15$ \\ 
\hline
\end{tabular}
\begin{list} {Table Notes.}
\item Note that EW of Mg$_{2}~\lambda5176$ is given in mag. Extraction area of all spectra of NGC\,6868
is $\rm3.66(\arcsec)^2\approx0.12\,kpc^2$. For NGC\,5903 it is $\rm4.57(\arcsec)^2\approx0.12\,kpc^2$, except 
for the outermost spectra, which is $\rm9.15(\arcsec)^2\approx0.25\,kpc^2$. EWs are corrected by velocity 
dispersion and the errors are the standard deviation of three different measurements of the EW for each line.
\end{list}
\end{table*}   

\begin{figure}
\resizebox{\hsize}{!}{\includegraphics[angle=0]{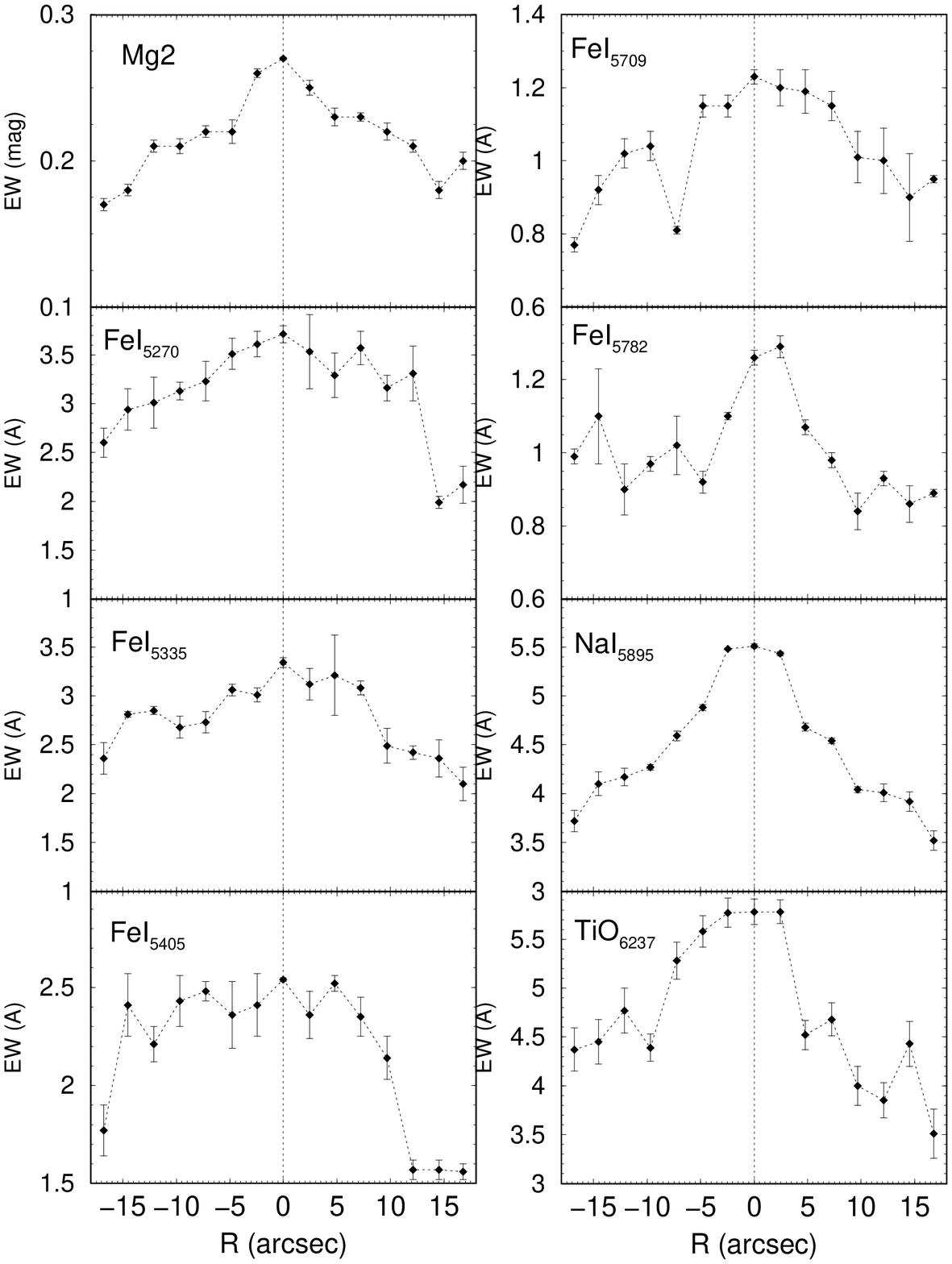}}
\caption{Spatial distribution of EWs measured in NGC\,6868.}
\label{fig5}
\end{figure}

\begin{figure}
\resizebox{\hsize}{!}{\includegraphics[angle=0]{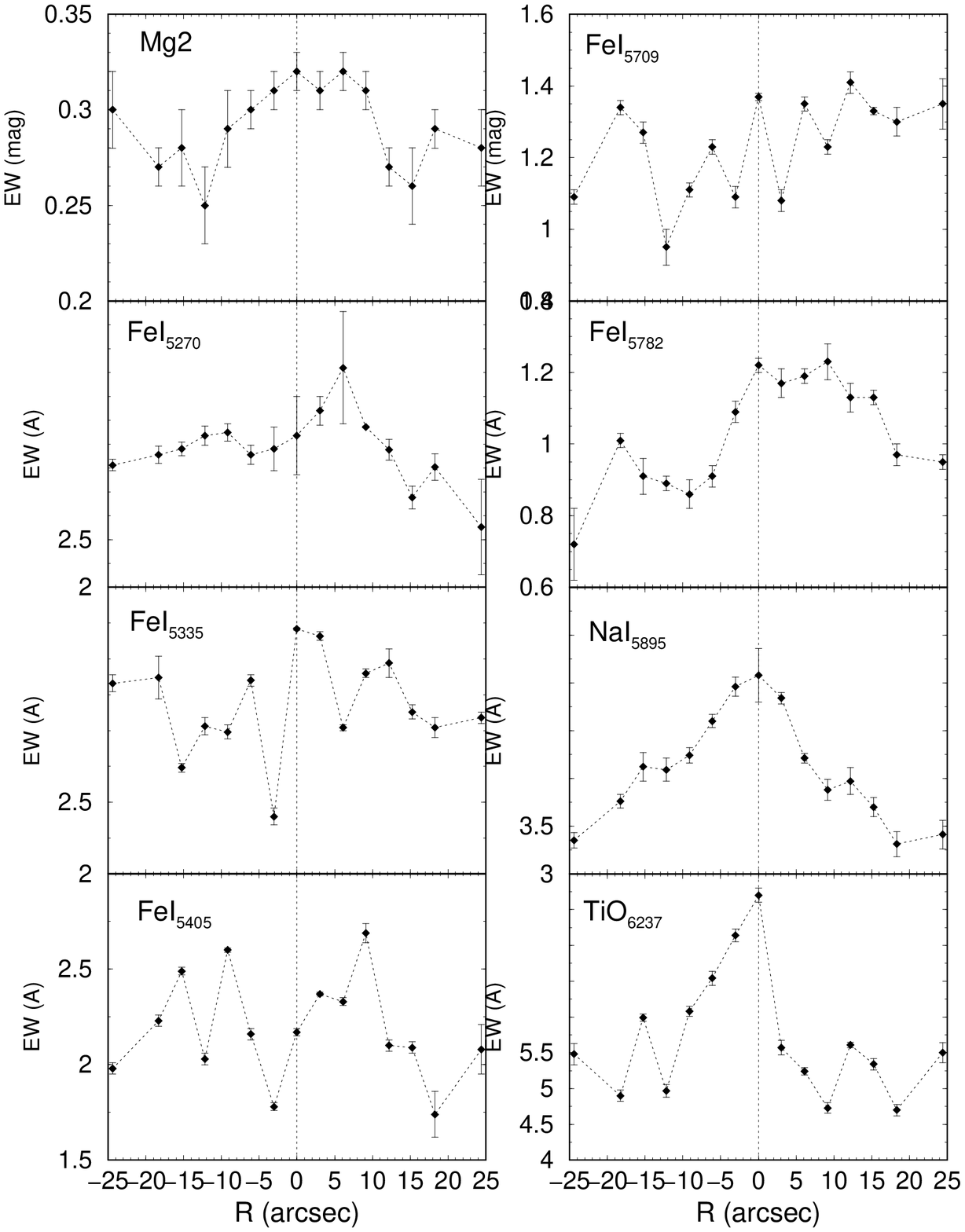}}
\caption{Same as Fig.~\ref{fig5} for the EWs of NGC\,5903.}
\label{fig6}
\end{figure}

The radial dependence of the EWs are shown in Figs.~\ref{fig5} and \ref{fig6}, respectively for
NGC\,6868 and NGC\,5903. Indices of NGC\,6868 consistently present a negative gradient 
(Fig.~\ref{fig5}) that indicates an overabundance of Fe, Mg, Na and TiO in the central parts with 
respect to the external regions.
   
A conspicuous negative gradient of the indices Mg$_{2}~\lambda5176$, ${\rm FeI}_{\lambda5270}$, 
${\rm FeI}_{\lambda5335}$, ${\rm NIa}_{\lambda5895}$ and ${\rm TiO}_{\lambda6237}$ is observed
in NGC\,5903 (Fig.~\ref{fig6}).

\begin{figure}
\resizebox{\hsize}{!}{\includegraphics[angle=0]{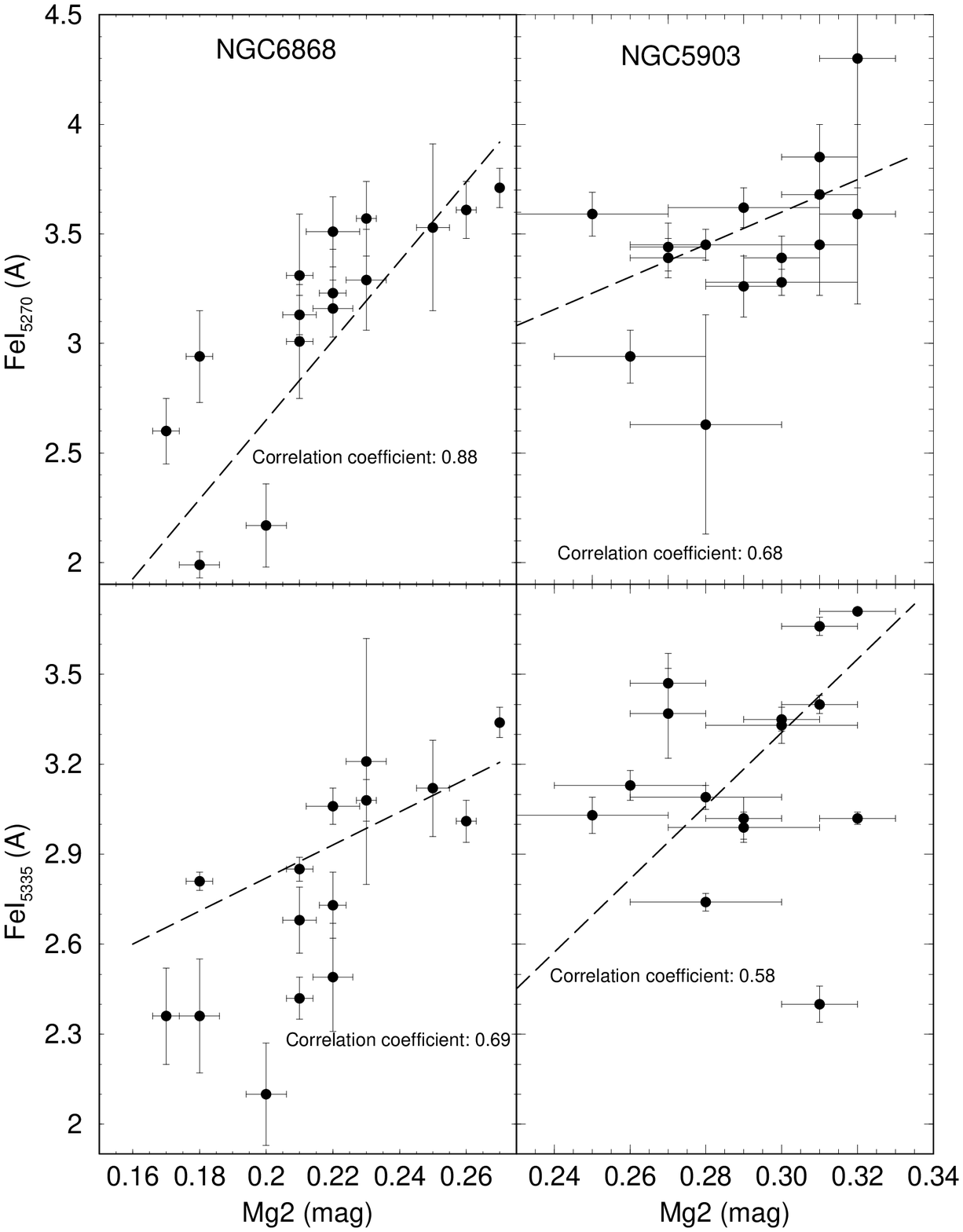}}
\caption{Correlations among selected Lick indices measured in NGC\,6868 (left panels) and
NGC\,5903 (right panels).}
\label{fig7}
\end{figure}

As shown in Fig.~\ref{fig7}, $Mg_2$ correlates both with ${\rm FeI}_{\lambda5270}$ and 
${\rm FeI}_{\lambda5335}$, which suggests that these elements probably underwent the
same enrichment process in NGC\,6868. However, only a marginal correlation
of $Mg_2$ and ${\rm FeI}_{\lambda5270}$ occurs in NGC\,5903.

If the galaxy gravitational potential is strong enough to retain the gas ejected by stars and
supernovae, it eventually migrates to the galaxy's central regions. New generations of stars
will be more metal rich at the center than in the external parts, and a steep radial negative
metallicity gradient is established as a mass-dependent parameter. Thus, a galaxy formed essentially
through a monolithic collapse is expected to present well-defined correlations between metallicity
gradients and mass \citep{car93a,ogan05}.

\citet{car93a} investigated the correlation of the gradient $d {\rm Mg_2}/dlog\,r$ with galaxy mass for a
sample of 42 elliptical galaxies, including NGC\,6868 and NGC\,5903. They found that the $Mg_2$ gradient
presents a bimodal trend with mass, in the sense that for galaxy mass smaller than $\sim10^{11}\ms$,
the gradient increases with mass. No pattern in the gradient was observed for larger masses. The increase
in the $Mg_2$ gradient with mass was interpreted as the result of a dissipative collapse associated with
star formation at work in the smaller galaxies. On the other hand, the lack of correlation between the
$Mg_2$ gradient with mass above the $\sim10^{11}\ms$ threshold might indicate that merging of smaller
galaxies could be the dominant formation mechanism of massive galaxies.

To test where NGC\,6868 and NGC\,5903 occur in the $d {\rm Mg_2}/dlog\,r$ {\em vs.} galaxy mass,
we compute their dynamical masses according to \citet{dokkum03}, with the data used in the present 
work. The results are $M_{NGC\,6868}=(3.2\pm0.1)\times10^{11}\,\ms$ and 
$M_{NGC\,5903}=(1.8\pm0.1)\times10^{11}\,\ms$, which agree, within the uncertainties, with the values 
computed by \citet{car93a}.
The average gradient values measured with our data are
$\rm (dMg_2/dlog\,r)_{NGC\,6868}=-0.07\pm0.01\,mag\,(\arcsec)^{-1}$ and
$\rm (dMg_2/dlog\,r)_{NGC\,5903}=-0.03\pm0.01\,mag\,(\arcsec)^{-1}$. Thus, according to the arguments
developed by \citet{car93a}, the present data are consistent with a merger origin for NGC\,6868 and
NGC\,5903.

\section{Stellar population synthesis}
\label{StellarPop}
 
The star formation history in a galaxy is a potential source of information 
not only on the age and metallicity distribution of the stellar population components.
Dating the star-forming episodes may provide clues to better understand galaxy formation and 
evolution processes.

In what follows we investigate the star formation history of NGC\,6868 and NGC\,5903 by means
of the stellar population synthesis method of \citet{bi88}. As population templates we use the
synthetic star cluster spectra of \citet{bruzual03}, built assuming a \citet{salpeter55}
Initial Mass Function and masses in the range $0.6 \leq m \leq 120\,\ms$.

Before establishing the spectral base, we note that elliptical galaxies, in general, do not present
recent star formation. Consequently, we only consider components older than 1\,Gyr in the synthesis.
The spectral base is made up of 7 components with ages of 1, 5 and 13\,Gyr and metallicities
$Z=0.008,\,\,0.02, \,\,{\rm and}\,\,0.05$ (Table~\ref{table_esquema}). These metallicities are 
taken as representative of the sub-solar, solar and above-solar ranges. 


\begin{table}
\renewcommand{\tabcolsep}{0.52cm}
\caption{Age and metallicity components}
\begin{tabular} {lccc}
\hline
      &$Z = 0.05$ & $Z = 0.02$ & $Z = 0.008$ \\
\hline
13\,Gyr & A1  & A2  & A3  \\
 5\,Gyr & B1  & B2  & B3  \\
 1\,Gyr &     & C2  &     \\
\hline
\end{tabular}
\label{table_esquema}
\end{table}

The stellar population synthesis provides directly the flux fractions (relative to the
flux at $\lambda 5870$) that each age and metallicity template contributes to the
observed spectrum. The sum of the template spectra, according to the individual flux
fractions, should be representative of the stellar population contribution to the
observed spectrum. A full description of the method is in \citet{rickes04}. 

\begin{figure}
\resizebox{\hsize}{!}{\includegraphics[angle=0]{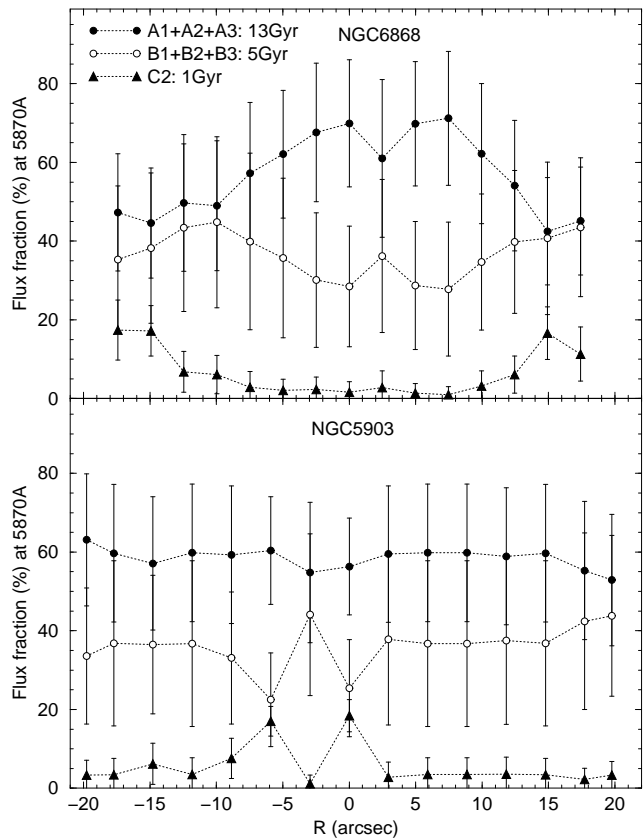}}
\caption{Flux-fraction contribution at $5870\,\AA$ of the stellar-population templates to the spectra
of NGC\,6868 (top panel) and NGC\,5903 (bottom panel). In both cases the 5 and 13\,Gyr 
components dominate the spectra, from the center to the external parts.}
\label{fig8}
\end{figure}

As a caveat we note that the observed spectra cover a relatively short spectral range and,
thus, contain a reduced number of indices sensitive to age and metallicity that, in principle, 
could result in a small number of constraints for the synthesis, especially in terms of metallicity.
However, the sum over metallicity shows that in the central regions of NGC\,6868 and NGC\,5903,
the 13\,Gyr population contributes with $70\pm16$\% and $56\pm12$\%, respectively. The 5\,Gyr
population contributes with $28\pm15$\% and $25\pm12$\%, respectively. These results suggest the 
presence of at least two populations of different ages in both galaxies, especially in the
central parts. The stellar population synthesis results are summarized in Fig.~\ref{fig8}, which
shows $\sim1\sigma$ differences in flux contribution between the 5 and 13\,Gyr populations,
along most of the galaxy's radial extent. Interestingly, the flux contribution of the 13\,Gyr 
population in NGC\,6868 increases towards the central region, while that of the 5\,Gyr decreases, 
within uncertainties. Flux fractions distribute almost symmetrically with respect to the
center of NGC\,6868. In NGC\,5903, both populations present almost uniform flux contributions, within 
uncertainties. The 1\,Gyr population presents significant flux contributions only in the external
regions ($|R|\ga2.3$\,kpc) of NGC\,6868, and in the central parts of NGC\,5903.

\begin{figure}
\resizebox{\hsize}{!}{\includegraphics[angle=0]{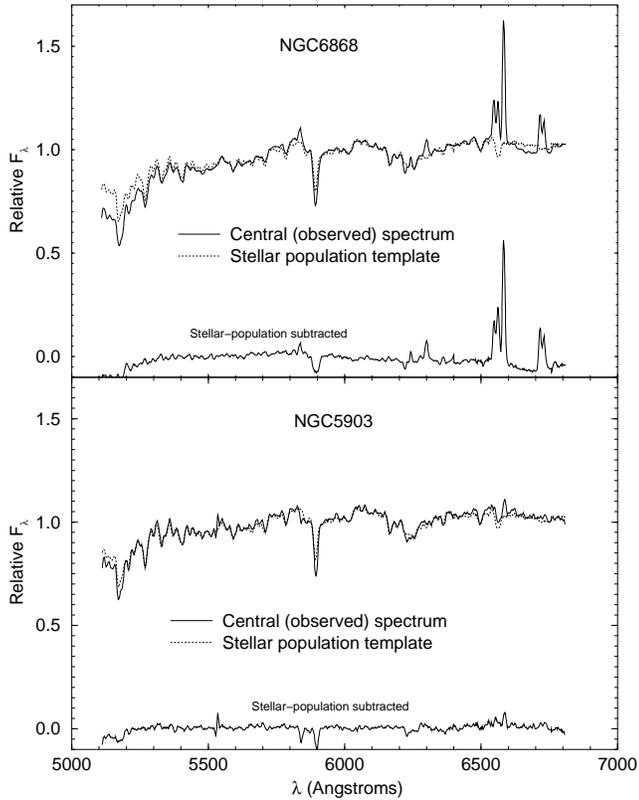}}
\caption{Stellar population synthesis of NGC\,6868 (top panel) and NGC\,5903 (bottom panel).
Notice that the subtraction of the stellar population enhanced $H\alpha$ in emission in 
NGC\,6868.}
\label{fig9}
\end{figure}

We illustrate the stellar population synthesis in Fig.~\ref{fig9}, where we 
superimpose on the central spectra of NGC\,6868 and NGC\,5903, the respective synthesized population
spectra. The residuals in the model fit to NGC\,6868 blueward of $\rm\sim 5500\,\AA$ are probably
due to bad calibration at the tail of the spectrum.
The stellar population-subtracted spectrum of NGC\,6868 present conspicuous emission lines,
especially Hydrogen Balmer, [NII] and [SII]. NGC\,5903, on the other hand, has no evidence of ionized
gas.

\section{Metallicity and ionized gas}
\label{MetGas}

\subsection{Metallicity}

Inferences on the metallicity of  NGC\,6868 and NGC\,5903 can be made comparing the observed Lick indices 
of Mg$_{2}~\lambda5176$, $FeI\,5270$ e $FeI\,5335$ with those derived from single-aged stellar population
(SSP) models (\citealt{thomas03}, \citealt{buzzoni94}), that assume a \citet{salpeter55} initial mass
function and 10\,Gyr of age.

\begin{figure}
\resizebox{\hsize}{!}{\includegraphics[angle=0]{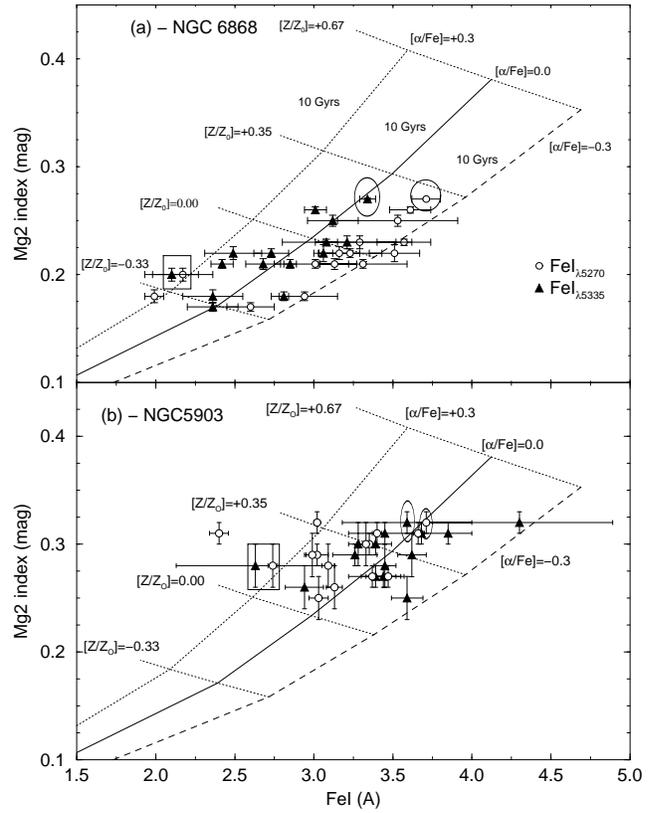}}
\caption{Selected Lick indices measured in NGC\,6868 (top panel) and NGC\,5903 (bottom panel)
are compared to those computed from SSP models for a range of metallicity and $\rm [\alpha/Fe]$ 
ratios. Ellipses and boxes indicate the central and external regions respectively.}
\label{fig10}
\end{figure}

Lick indices computed from SSP models for different metallicities and $\rm [\alpha/Fe]$ ratios together
with our data  are shown in Fig.~\ref{fig10}, with NGC\,6868 in the top panel and NGC\,5903 in the bottom.
The central parts of NGC\,6868 ($|R|\la0.5$\,kpc) present a deficiency of alpha elements with respect to
iron ($\rm-0.3\la[\alpha/Fe]\la0.0$) and an above-solar metallicity ($[Z/Z_\odot]\approx+0.3$. The external
parts, on the other hand, present a higher, roughly uniform distribution of ratios
($\rm [\alpha/Fe]\approx+0.3$) and sub-solar metallicities ($[Z/Z_\odot]\approx-0.33$). The overabundance
of the $\rm [\alpha/Fe]$ ratios has been interpreted as a consequence of the chemical enrichment produced
by type II supernovae with respect to type Ia ones (\citealt{Idiart03}). Thus, differences measured in
$\rm [\alpha/Fe]$ imply different star-formation histories for the central and external regions of
NGC\,6868, which is consistent with the results derived from the stellar population synthesis
(Sect.~\ref{StellarPop}). A similar conclusion applies to NGC\,5903 (bottom panel), since the central
parts have metallicities in excess of $[Z/Z_\odot]\approx+0.35$ and $\rm [\alpha/Fe]\approx0.0$, while the
external parts have metallicities between solar and $[Z/Z_\odot]\approx+0.35$ and present an enhanced ratio
$\rm [\alpha/Fe]\approx+0.3$.



\subsection{Ionized gas}
\label{IonizedGas}

Emission gas has been recently detected in a large number of elliptical galaxies \citep{phi86}. However,
its origin and the nature of the ionization source have not yet been conclusively established. NGC\,6868
presents conspicuous emission lines not only in the central region, but in spectra extracted up to
$\rm R\sim17\arcsec\sim3.1\,kpc$ (Fig.~\ref{fig3}). In what follows we investigate properties of the emission
gas using fluxes of the lines $H\alpha$, $[NII]_{\lambda\lambda\,6548,6584}$, $[OI]_{\lambda\,6300}$ and
$[SII]_{\lambda\lambda\,6717,6731}$, measured in the stellar population-free spectra, i.e., those resulting
from the subtraction of the respective population templates (Sect.~\ref{StellarPop}). The emission line fluxes
were measured fitting Gaussian to the profiles.

\begin{figure}
\resizebox{\hsize}{!}{\includegraphics[angle=0]{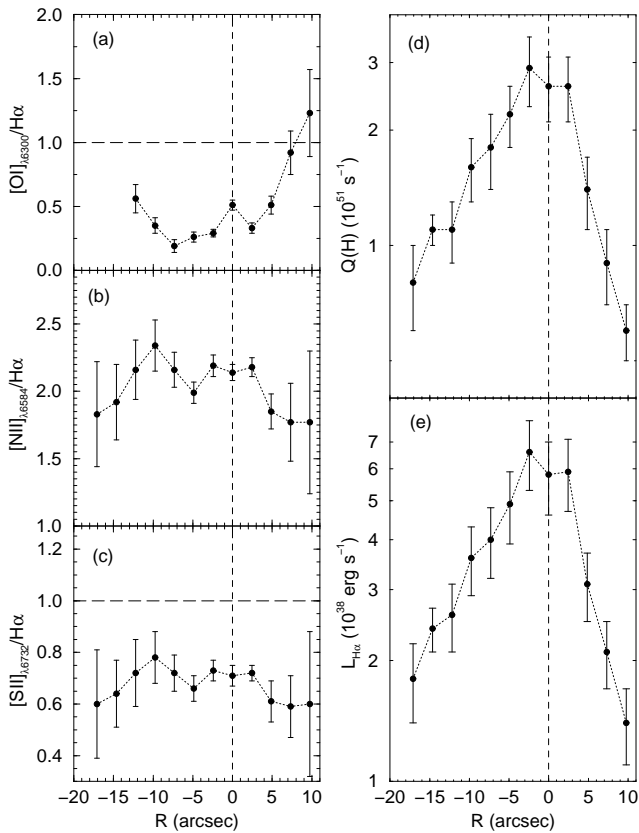}}
\caption{Spatial variation of the ratios $\frac{[OI]_{\lambda\,6300}}{H\alpha}$ 
(panel a), $\frac{[NII]_{\lambda6584}}{H\alpha}$ (b) and $\frac{[SII]_{\lambda6731}}{H\alpha}$
(c). Panels (d) and (e) show the spatial distribution of the number of ionization photons
and $H\alpha$ luminosity, respectively. The dashed line in panels (a) and (c) discriminates
between different ionizing sources (Sect.~\ref{PAGB}).}
\label{fig11}
\end{figure}

Flux ratios with respect to $H\alpha$ are given in Tab.~\ref{tableNIISII}, and the 
spatial distribution of the measured ratios $\frac{[OI]}{H\alpha}$, $\frac{[NII]}{H\alpha}$ and
$\frac{[SII]}{H\alpha}$ are shown in the left panels of Fig.~\ref{fig11}.
    
\begin{table}
\renewcommand{\tabcolsep}{0.18cm}
\renewcommand{\arraystretch}{1.2}
\caption{Emission line parameters of NGC\,6868}
\label{tableNIISII}
\begin{tabular}{lcccc}
\hline
$R(\arcsec)$&$H\alpha$&$\frac{[NII]}{H\alpha}$& $\frac{[SII]}{H\alpha}$ & $\frac{[OI]}{H\alpha}$\\
(1)&(2)&(3)&(4)&(5)\\
\hline
0.00   & $3.36\pm0.08$ &  $2.81\pm0.21$ & $1.23\pm0.10$ & $0.51\pm0.04$ \\
2.44S  & $3.39\pm0.10$ &  $2.90\pm0.91$ & $1.18\pm0.12$ & $0.33\pm0.04$ \\
4.88S  & $1.81\pm0.11$ &  $2.46\pm0.21$ & $0.92\pm0.11$ & $0.51\pm0.07$ \\
7.32S  & $1.20\pm0.18$ &  $2.36\pm0.36$ & $0.88\pm0.40$ & $0.92\pm0.17$ \\
9.76S  & $0.82\pm0.21$ &  $2.36\pm0.61$ & $1.20\pm0.31$ & $1.23\pm0.34$ \\
\hline
2.44N  & $3.83\pm0.13$ &  $2.92\pm0.32$ & $1.32\pm0.11$ & $0.29\pm0.03$ \\
4.88N  & $2.86\pm0.10$ &  $2.65\pm0.22$ & $1.30\pm0.12$ & $0.26\pm0.04$ \\
7.32N  & $2.34\pm0.13$ &  $2.87\pm0.64$ & $1.25\pm0.10$ & $0.19\pm0.05$ \\
9.76N  & $2.08\pm0.16$ &  $3.11\pm0.42$ & $1.39\pm0.11$ & $0.35\pm0.06$ \\
12.20N & $1.49\pm0.13$ &  $2.88\pm0.33$ & $1.47\pm0.41$ & $0.56\pm0.11$ \\
14.64N & $1.39\pm0.18$ &  $2.56\pm0.57$ & $1.23\pm0.17$ & $-$\\
17.08N & $1.06\pm0.20$ &  $2.43\pm0.48$ & $1.20\pm0.23$ & $-$\\
\hline
\end{tabular}
\begin{list} {Table Notes.}
\item $H\alpha$ flux in col.~2 is given in $\rm 10^{-15}\,erg\,s^{-1}\,cm^{-2}$. 
\end{list}
\end{table}
 
Assuming case B recombination, the number of ionizing photons can be computed from $H_\alpha$
luminosity using the equation $Q(H) =\frac{L_{H\alpha}}{h\nu_\alpha}\frac{\alpha_B
(H^0,T)}{\alpha_{H\alpha} (H^0,T)}$ \citep{oster89}, where $\alpha_B\,(H^0,T)$ is the
total recombination coefficient, and $\alpha_{H\alpha}\,(H^0,T)$ is the recombination coefficient
for ${\rm H}\alpha$. Values of $Q(H)$ and $L_{H\alpha}$ for NGC\,6868 are given in
Tab.~\ref{luminosidade}, and their spatial distributions are shown in panels (d) and (e) of
Fig.~\ref{fig11}. 

\begin{table}
\renewcommand{\tabcolsep}{0.75cm}
\caption{$H\alpha$ luminosity and number of ionizing photons in NGC\,6868}
\label{luminosidade}
\begin{tabular}{ccc}
\hline
$R$ & $L_{H_\alpha}$ & $Q(H)$ \\
(\arcsec)& $(\rm 10^{38}\,erg\,s^{-1})$ & $(\rm 10^{51}\,s^{-1})$ \\
\hline
0.00    & $5.8\pm1.2$ & $2.6\pm0.5$  \\
2.44S   & $5.9\pm1.2$ & $2.6\pm0.5$  \\
4.88S   & $3.1\pm0.6$ & $1.4\pm0.3$  \\
7.32S   & $2.1\pm0.4$ & $0.9\pm0.2$  \\ 
9.76S   & $1.4\pm0.3$ & $0.6\pm0.1$  \\
\hline           
2.44N   & $6.6\pm1.3$ & $2.9\pm0.6$  \\
4.88N   & $4.9\pm1.0$ & $2.2\pm0.4$  \\
7.32N   & $4.0\pm0.8$ & $1.8\pm0.4$  \\
9.76N   & $3.6\pm0.7$ & $1.6\pm0.3$  \\
12.20N  & $2.6\pm0.5$ & $1.1\pm0.2$  \\
14.64N  & $2.4\pm0.3$ & $1.1\pm0.1$  \\
17.08N  & $1.8\pm0.4$ & $0.8\pm0.2$  \\
\hline
\end{tabular}
\begin{list} {Table Notes.}
\item $H\alpha$ luminosity and number of ionizing photons computed for the
extractions given in col.~1 of NGC\,6868.
\end{list}
\end{table}

Fig.~\ref{fig11} also shows that in all extracted spectra of NGC\,6868, the ratio 
$\frac{[NII]_{\lambda6584}}{H\alpha}$ is larger than 1 and $\frac{[SII]_{\lambda6731}}{H\alpha}$ 
is smaller than 1, within uncertainties.

With the above facts in mind we consider three possible scenarios to infer the nature
of the ionization source in NGC\,6868. They involve the presence of star clusters,
post-AGB stars and an AGN.

\subsubsection{Star cluster}

To test this hypothesis we build a star cluster that produces a total number of ionizing photons of
the order of $\rm 10^{51}\,s^{-1}$ following \citet{salpeter55} initial mass function,
$\phi(m)\propto m^{-(1+\chi)}$, with $\chi=1.35$, considering stars in the mass range $0.1 - 100\,\ms$. 
For the
relation of number of ionizing photons ($N_{Ly}$) with stellar mass we use the library of stellar
atmospheres of \citet{kur79}. In cols.~2 and 3 of Table~\ref{estrelas} we provide $N_{Ly}$ for ionizing
stars in the mass range $10 - 30\,\ms$. In this mass range, the number of ionizing photons can be
related to stellar mass by the approximation
$N_{Ly}(m)\approx3.9\times10^{39}\left(\frac{m}{\ms}\right)^{5.9}\rm s^{-1}$. Thus, the total number of
ionizing photons can be computed from $\int_{10}^{30}N_{Ly}(m)\phi(m) dm$.

The resulting distribution of stars in terms of spectral type of the hypothetical ionizing star cluster
is given in col.~4 of Tab.~\ref{estrelas} for representative spectral types. Because of the large number
of O5 stars, such a star cluster would be very young, luminous and have a mass of $\sim10^6\,\ms$.

\begin{table}
\caption{Hypothetical ionizing star cluster}
\begin{tabular}{cccc}
\hline
Spectral type&$m$& $N_{Ly}$& Number of stars \\
 &(\ms) & ($\rm 10^{48}\,s^{-1}$) \\
 (1) & (2) & (3) & (4)\\
\hline
O5  &30 & 2.00  &226 \\
    &22 & 0.25  &470 \\
B0  &17 & 0.082 &860 \\
    &15 & 0.043 &1154 \\
    &13 & 0.015 &1616 \\
    &10 & 0.0021&2993 \\
\hline
\end{tabular}
\label{estrelas}
\end{table}

We conclude that ionization by such a massive star cluster can be ruled out, because with more than 200 O5
stars it should be bright enough to be easily identified in images. Besides, its flux contribution to the
observed spectra should be conspicuously detected by the stellar population synthesis (Sect.~\ref{StellarPop}).

\subsubsection{Post-AGB stars}
\label{PAGB}

Since a typical Post-AGB star emits $\rm\sim10^{47}\,s^{-1}$ \citep{bine94}, about $10^4$ such stars would
be necessary to produce the amount of ionizing photons measured in each extracted region (Tab.~\ref{estrelas}).
All spectra of NGC\,6868 were extracted from regions with an area of $\approx0.19\,\rm kpc^2$. This implies a 
projected number density of Post-AGB stars of $\sim8\times10^4\,\rm kpc^{-2}$ scattered throughout the
$R\approx3.3$\,kpc central region. Such a density is not absurd \citep{bine94}, however, Post-AGB star
ionization models for a variety of ionization parameters predict that the intensity ratio
$[OI]_{\lambda6300}/H\alpha$ should be smaller than unity, while those of $[SII]_{\lambda6731}/H\alpha$ 
and $[NII]_{\lambda6584}/H\alpha$ should be smaller than $\approx1.5$. The first two conditions are met
by the measurements of NGC\,6868, but the ratios $[NII]_{\lambda6584}/H\alpha$ are all above 1.5 
(Fig~\ref{fig11}).
     
The presence of a significant number of post-AGB stars in early-type galaxies produces a
conspicuous excess in the UV flux, e.g. \citet{Bica96}. NGC\,6868 is included in the galaxies
investigated by \citet{Bonatto96}, among those with a flat UV spectrum, which is consistent with
the above arguments against a numerous population of post-AGB stars in this galaxy.

\subsubsection{Active Nucleus}

To test this hypothesis we build diagnostic-diagrams involving the ratios $[NII]_{\lambda6584}/H\alpha$,
$[OI]_{\lambda6300}/H\alpha$ and $[SII]_{\lambda6717,31}/H\alpha$, using as comparison the values measured
in a sample if nearby galaxies with low luminosity ($\rm L_{H\alpha}<2\times10^{39}\,erg\,s^{-1}$)
active nucleus (\citealt{ho97}). Here we use the galaxies in \citet{ho97} sample with unambiguous classification
as LINER in spiral, LINER in elliptical, Seyfert or Starburst. The restricted sample amounts to 128 galaxies.

\begin{figure}
\resizebox{\hsize}{!}{\includegraphics[angle=0]{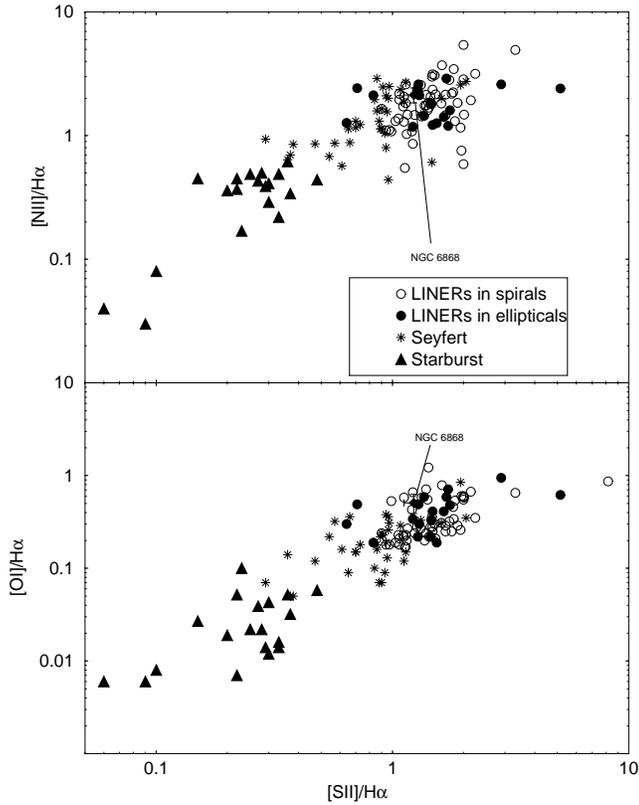}}
\caption{Emission-line diagnostic diagrams with a subsample of the galaxies in \citet{ho97} plotted
as comparison. The locus of NGC\,6868 in both diagrams is consistent with that of LINERs. }
\label{fig12}
\end{figure}

The results are shown in Fig.~\ref{fig12}, where in both panels we see that the values measured in NGC\,6868
consistently fall in the locus occupied predominantly by LINERs.

These results, together with those of the stellar population synthesis, suggest that the main source 
of gas ionization in NGC\,6868 is non-thermal. However, as a caveat we note that shocks can also 
produce emission-line ratios similar to those measured in NGC\,6868. In this sense, the  ratios $[NII]_{\lambda6584}/H\alpha$ and $[SII]_{\lambda6717,31}/H\alpha$ alone cannot discriminate between 
the AGN and shocks as the ionization source. Indeed, \citet{thaisa96} show that a low-luminosity AGN
cannot ionize gas up to distances of $\sim3.5$\,kpc. Besides, they suggest that ionization in the 
external region is produced by shocks, which accounts for the constancy of the $[NII]_{\lambda6584}/H\alpha$ 
and $[SII]_{\lambda6717,31}/H\alpha$ ratios.

\section{Discussion and concluding remarks}
\label{Dicussion}

The stellar population, metallicity distribution and ionized gas in NGC\,6868 and NGC\,5903 have
been investigated in this paper by means of long-slit spectroscopy and stellar population synthesis.
Lick indices  of both galaxies  consistently present a negative gradient that indicates an overabundance
of Fe, Mg, Na and TiO in the central parts with respect to the external regions. We found that $Mg_2$
correlates both with ${\rm FeI}_{\lambda5270}$ and ${\rm FeI}_{\lambda5335}$, which suggests that these
elements probably underwent the same enrichment process in NGC\,6868. However, in NGC\,5903 only a marginal
correlation of $Mg_2$ and ${\rm FeI}_{\lambda5270}$ occurs. 

Galaxy mass and the $Mg_2$ gradient computed in the present work are consistent with previous results 
that suggest that NGC\,6868 and NGC\,5903 were formed by merger events. In addition, the stellar population
synthesis clearly shows the presence of at least two populations of different ages in both galaxies.
Particularly in the central regions of NGC\,6868 and NGC\,5903, the 13\,Gyr population contributes with
$70\pm16$\% and $56\pm12$\%, respectively. The 5\,Gyr population contributes with $28\pm15$\% and
$25\pm12$\%, respectively in flux fraction at $\rm5870\,\AA$. 

The central regions of NGC\,6868 and NGC\,5903 present a deficiency of alpha elements with respect 
to the external parts, which suggests different star-formation histories in both regions. With 
respect to the metallicity, the central regions of both galaxies have higher $[Z/Z_\odot]$ values
than the external parts. The range in metallicity spanned by the sampled regions of NGC\,6868 appears 
to be larger than in NGC\,5903.

Concerning the ionized gas in NGC\,6868, the ratios $[NII]_{\lambda6584}/H\alpha$, $[OI]_{\lambda6300}/H\alpha$
and $[SII]_{\lambda6717,31}/H\alpha$ consistently suggest the presence of a LINER at the galaxy center.
These results, together with the stellar population synthesis, suggest that the main source of gas ionization
in NGC\,6868 is non-thermal, produced by a low-luminosity active nucleus, probably with some contribution
of shocks to explain ionization at distances of $\sim3.5$\,kpc from the nucleus.

\section*{Acknowledgments}
We thank the anonymous referee for helpful suggestions.
We acknowledge the Brazilian agency CNPq for partial support of this work. This research has made use of
the NASA/IPAC Extragalactic Database (NED) which is operated by the Jet Propulsion Laboratory, California
Institute of Technology, under contract with the National Aeronautics and Space Administration.


\begin{thebibliography}{}


\bibitem[\protect\citeauthoryear{Arimoto \& Yoshii }{1987}]{arim87}
   Arimoto, N., YosH\,II, Y. 1987, A\&A, 23, 38

\bibitem[\protect\citeauthoryear{Bica}{1988}]{bi88}
   Bica, E. 1988, A\&A, 195, 79

\bibitem[\protect\citeauthoryear{Bica et al.}{1996}]{Bica96}
   Bica, E., Bonatto, C., Pastoriza, M.G. \& Alloin, D. 1996, A\&A, 313, 405

\bibitem[\protect\citeauthoryear{Bica}{1988}]{bi288}
   Bica, E., Arimoto, N. \& Alloin, D. 1988, A\&A, 202, 8

\bibitem[\protect\citeauthoryear{Binette}{1994}]{bine94}
   Binette, L. Magris, C. G. Stasinska, G. Bruzual, A.G., 1994, A\&A, 292, 13

\bibitem[\protect\citeauthoryear{Bonatto et al.}{1996}]{Bonatto96}
   Bonatto, C., Bica, E., Pastoriza, M.G. \& Alloin, D., 1996, A\&A, 118, 89

\bibitem[\protect\citeauthoryear{Bruzual \& Charlot}{2003}]{bruzual03}
   Bruzual, G., Charlot, S. 2003, MNRAS, 344, 1000

\bibitem[\protect\citeauthoryear{Burkert \& Naab}{2000}]{burk00}
   Burkert, A.; Naab, T. 2000, A\&AS, 23, 1497

\bibitem[\protect\citeauthoryear{Buzzoni et al.}{1992}]{buzzoni94}
   Buzzoni, A.; Mantegazza, L., Gariboldi, G. 1994, ApJ., 107, 513

\bibitem[\protect\citeauthoryear{Caon et al.}{2000}]{ca00}
   Caon, N., Macchetto, D., Pastoriza, M. 2000, ApJS, 127, 39
   
\bibitem[\protect\citeauthoryear{Cardelli et al.}{1989}]{CCM89}
   Cardelli, J. A., Clayton, G. C. \& Mathis, J. S. 1989, ApJ, 345, 245

\bibitem[\protect\citeauthoryear{Carollo et al.}{1993a}]{car93a}
   Carollo, C. M., Danziger, I.J., Buson, L. 1993a, MNRAS, 265, 553

\bibitem[\protect\citeauthoryear{Carollo et al.}{1993b}]{car93b}
   Carollo, C. M., Danziger, I.J., Buson, L. 1993b, MNRAS, 553, 580

\bibitem[\protect\citeauthoryear{Carollo et al.}{1994a}]{car94a}
   Carollo, C. M., Danziger, I.J. 1994a, MNRAS, 270, 523

\bibitem[\protect\citeauthoryear{Carollo et al.}{1994b}]{car94b}
   Carollo, C. M., Danziger, I.J. 1994b, MNRAS, 270, 743

\bibitem[\protect\citeauthoryear{Chiosi \& Carraro}{2002}]{ChioCar02}
   Chiosi, C. \& Carraro, G. 2002, MNRAS, 335, 335

\bibitem[\protect\citeauthoryear{Davis et al.}{1987}]{davis87}
   Davies, R.L., Burstein, D., Dressler, A., Faber, S.M., Lynden-Bell, D.,
   Terlevich, R., Wegner, G. 1987, ApJS, 64, 581

\bibitem[\protect\citeauthoryear{Davis et al.}{1993}]{davis93}
   Davies, R.L., Sadler, E.M., Peletier, R.F. 1993, MNRAS, 262, 650

\bibitem[\protect\citeauthoryear{Davis et al.}{1993}]{davis93b}
   Davies, R.L., Sadler, E.M., Peletier, R.F. 1993, MNRAS, 650, 680

\bibitem[\protect\citeauthoryear{de Vaucouleurs et al.}{1991}]{vaucouleurs}
   de Vaucouleurs, G., de Vaucouleurs, A., Corwin, H. G., Jr., Buta, R. J.,
   Paturel, G., Fouque, P., 1991.

\bibitem[\protect\citeauthoryear{Faber et al.}{1985}]{fab85}
   Faber, S. M., Friel, E. D., Burstein, D., Gaskell, C. M. 1985, ApJS, 57, 711

\bibitem[\protect\citeauthoryear{Ferrari et al.}{1999}]{ferr99}
   Ferrari, F., Pastoriza, M. G., Macchetto, F., Caon, N. 1999, A\&A, 136, 269

\bibitem[\protect\citeauthoryear{Ferrari et al.}{2002}]{ferr02}
   Ferrari, F.; Pastoriza, M. G.; Macchetto, F. D.; Bonatto, C.; Panagia, N.;
   Sparks, W. B. 2002, A\&A, 389, 355

\bibitem[\protect\citeauthoryear{Ho et al.}{1997}]{ho97}
   Ho, L.C., Filippenko, A.V., Sargent, W.L.W., 1997, APJ, 112, 315

\bibitem[\protect\citeauthoryear{Idiart et al.}{2003}]{Idiart03}
   Idiart, T.P., Michard, R. \& de Freitas Pacheco, J.A., 2003, A\&A, 398, 949

\bibitem[\protect\citeauthoryear{Kurucz}{1979}]{kur79}
   Kurucz, R. L. 1979, ApJS, 40, 1

\bibitem[\protect\citeauthoryear{Macchetto et al.}{1996}]{macc96}
   Macchetto, F., Pastoriza, M., Caon, N., Sparks, W.B., Giavalisco, M.,
   Bender, R., Capaccioli, M. 1996, A\&A, 120, 463

\bibitem[\protect\citeauthoryear{Maia et al.}{1989}]{maia89}
   Maia, M. A. G., Da Costa, L. N. Latham, D. W., 1989, AJSS, 69, 809

\bibitem[\protect\citeauthoryear{Ogando et al.}{2005}]{ogan05}
   Ogando, R.L.C., Maia, M.A.G., Chiappini, C., Pellegrini, P.S., Schiavon, R.P.,
   da Costa, L.N.  2005, ApJ, 61, 64

\bibitem[\protect\citeauthoryear{O'Sullivan et al.}{2001}]{osull01}
   O`Sullivan, E., Forbes, D. A., Ponman, T. J., 2001, MNRAS, 328, 461

\bibitem[\protect\citeauthoryear{Osterbrock}{1989}]{oster89}
   Osterbrock, D. E., 1989, in "Astrophisics of Gaseous Nebulae and Active Galactic Nuclei", 
University Science Books

\bibitem[\protect\citeauthoryear{Phillips et al.}{1986}]{phi86}
   Phillips, M. M., Jenkins, C. R., Dopita, M. A., Sadler, E. M.,
   Binette, L., 1986, ApJ, 91, 1062

\bibitem[\protect\citeauthoryear{Rickes et al.}{2004}]{rickes04}
   Rickes, M. G,; Pastoriza, M.G., Bonato, C., 2004, 419, 449

\bibitem[\protect\citeauthoryear{Salpeter}{1955}]{salpeter55}
   Salpeter, E.F., 1979, ApJ, 121, 161

\bibitem[\protect\citeauthoryear{Savage et al.}{1977}]{sava77}
   Savage, A., Wright, A. E., Bolton, J. G. 1977, AuJPAS, 44, 1

\bibitem[\protect\citeauthoryear{Schweizer et al.}{1990}]{schw90}
   Schweizer, F., Seitzer, P., Faber, S. M.; Burstein, D., Dalle Ore,
   C. M., Gonzalez, J. J. 1990, ApJ, 364, 33

\bibitem[\protect\citeauthoryear{Sparks et al.}{1985}]{spar85}
   Sparks, W.B., Wall, J.V., Thorne, D.J., Jorden, P.R, van Breda, I.G.,
   Rudd, P.J., Jorgensen, H.E., 1985, MNRAS, 217, 89

\bibitem[\protect\citeauthoryear{van Dokkum \& Stanford}{2003}]{dokkum03}
   van Dokkum, P., Standord, S. A., 2003, ApJ, 285, 79

\bibitem[\protect\citeauthoryear{Storchi-Bergmann et al.}{1996}]{thaisa96}
   Storchi-Bergmann, Thaisa; Wilson, Andrew S.; Baldwin, Jack A., 1996, APJ, 460, 252

\bibitem[\protect\citeauthoryear{Thomas et al.}{2003}]{thomas03}
   Thomas, D., Maraston, C., Bender, R., 2003, MNRAS, 339, 897

\bibitem[\protect\citeauthoryear{Worthey}{1994}]{wor94}
   Worthey, G. 1994, ApJS, 95, 107

\bibitem[\protect\citeauthoryear{Worthey et al.}{1992}]{wor92}
   Worthey, G.; Faber, S. M.; Gonz\'ales, J. J . 1992, ApJS, 398, 69

\bibitem[\protect\citeauthoryear{Zeilinger et al.}{1996}]{zeil96}
   Zeilinger, W.W., Pizzella, A., Amico, P., Bertin, G., Bertola, F.,
   Buson, L. M., Danziger, I. J., Dejonghe, H., Sadler, E. M., Saglia, R. P.,
   de Zeeuw, P. T., 1996, A\&AS, 120, 257



\end{thebibliography}
\end{document}